\documentclass[twocolumn,amsmath,amssymb]{revtex4} 
\usepackage{graphicx}
\begin{document} 
\title{Scalable Ion Trap Quantum Computation with Pairwise Interactions Only} 
\author{K. R. Brown, J. Vala, and K. B. Whaley} 
\affiliation{Department of Chemistry, Univesity of California, Berkeley, California, 94720} 
\date{\today } 
 
\begin{abstract} 
Universal ion trap computation on Decoherence Free Subspaces (DFS) using only two qubit operations is presented.  
The DFS is constructed for the collective dephasing model. 
Encoded single and two-qubit logical operations are implemented via the Sorensen-Molmer interaction. 
Alternation of the effective Hamiltonians for two particular phase configurations of control fields approximates an anisotropic exchange interaction. 
This is universal over suitable encodings of one logical qubit into three physical qubits which are also DFS under collective decoherence. 
\end{abstract} 
\maketitle 
 
\section{Introduction} 
 
Quantum computation is capable of cracking some hard problems of both classical computation and quantum physics \cite{Nielsen:00}.  
However, it meets non-trivial difficulties on its way to physical implementation \cite{DiVincenzo:95a}. 
One persisting source of troubles is decoherence, which smears out the results of the quantum computation, rendering them useless.  
One promising physical implementation for quantum computation that possesses relatively low decoherence is using trapped ions \cite{Wineland:98,Cirac:95}. 
In this paper, we present an approach for doing quantum computation with trapped ions that allows one to use only pairwise interaction and stay in a Decoherence Free Subspace (DFS) \cite{Palma:96,Duan:97,Zanardi:97a,Lidar:98}.    
DFS are a subspace of states whose symmetry makes particular error operators act like the identity on the subspace.    
 
For ion traps, an important source of errors is long wavelength fluctuating magnetic fields.   
For neighboring groups of ions such long wavelength magnetic fields act as a collective phase error.   
A DFS exists for the collective phase error which has been studied by several groups. \cite{Palma:96,Zanardi:97a,Kempe:01a}.   
Experiments have demonstrated use of this for protection of two-qubit \cite{Kwiat:00,Kielpinski:01,Fortunato:02} and three-qubit systems \cite{Viola:01}.
Experimentally, Kielpinski et al. \cite{Kielpinski:01} demonstrated that these DFS are important in the context of ion traps, showing an increased phase coherence when one logical qubit is encoded into the DFS that is constructed from two physical qubits. 
Recently, Kielpinski et al. \cite{Kielpinski:02} have developed a method to perform computation on this DFS uses entangling operations between groups of both two and four physical qubits. A method to compute on a different DFS, one that protects the ion trap quantum computer from errors due to spontaneous emission, was devolped in Ref. \cite{Beige:00}. 

In this paper, we show a different way to perform universal computation on a collective dephasing DFS, using only entangling operations between two physical qubits. 
We explain how this could be done in linear ion traps using operations that have already been established.   
In Section \ref{ion traps} we briefly describe the physical system and then show how one can do universal computation using certain prepared encoded states and two Sorensen-Molmer-type gates \cite{Sorensen:99,Nielsen:00}.   
In Section \ref{encoded}, we will explain both why these gates are universal and why the encoded subspace is a DFS.  
In Section \ref{end}, we conclude with some remarks about the feasibility of such an approach compared to other schemes for universal quantum computation in ion traps. 

\section{Effective Exchange Interactions in Ion Traps} \label{ion traps} 
  
Our system is a string of $N$ two level ions in a linear ion trap \cite{Cirac:95,Wineland:98}.  
In a linear trap the confinement in the radial direction is stronger than the 
confinement in the axial direction. For small displacements of the atoms 
one can treat the motion of the system as N coupled one-dimensional harmonic 
oscillators. Detailed calculations of the normal modes for a large number of 
atoms in experimentally realizable traps are now widely available  \cite{James:98}. 
The scheme we outline below  requires that each atom be individually 
addresed by a control laser. 
 
Coupling between atoms is brought about by coupling the internal state of the ions to 
the motional mode of the string of ions.  A variety of gates have been 
suggested to achieve this coupling \cite{Cirac:95,Sorensen:99,Childs:00,Knight:00}. 
For our purposes, the coupling gate introduced by Sorensen and Molmer \cite{Sorensen:99} proves most useful. 
  
The Sorensen-Molmer gate, described in \cite{Sorensen:99}, couples two ions via a two photon process that 
virtually populates the excited motional modes of the ion string.  
We briefly summarize their scheme here.  
By  applying two lasers of opposite detuning to two ions, $i$ and $j$, one 
creates the following effective Hamiltonian: 
 
\begin{equation} 
H_{1}=\frac{\eta ^{2}\Omega ^{2}}{\Delta }\sigma _{x}^{i}\sigma _{x}^{j} 
\end{equation} 
where $\Omega $ is the on-resonant Rabi frequency of each atom, $\eta $ is 
the Lamb-Dicke parameter, $\Delta $ is the detuning, and $\sigma _{x}^{i}$ 
is the Pauli spin operator $x$ on atom $i$.  The Lamb-Dicke parameter $\eta $ is 
defined as the ration of the width of the zero point motion of the ion 
divided by the wavelength of applied field.  In terms of the ion trap 
parameters, $\eta =\sqrt{\frac{\hbar}{2NM\omega }}\frac{1}{\lambda }\cos 
\theta $, where $\omega $ is the frequency of the trap, $M$ is the mass of 
the ion, $\lambda $ is the wavelength of the driving field, and $\theta $ is 
the angle between the wavevector of the light and the axial direction of the 
trap \cite{Wineland:98}.  
 
By changing the phase of the incoming laser pulses by $\pi /2$, one can 
apply the slightly modified Hamiltonian, 
 
\begin{equation} 
H_{2}=\frac{\eta ^{2}\Omega ^{2}}{\Delta }\sigma _{y}^{i}\sigma _{y}^{j} 
\end{equation} 
Note that $H_{1}$and $H_{2}$ commute. Therefore, by applying $H_{1}$ followed by $H_{2}$ for equal times, one can perform the following unitary 
transformation 
 
\begin{equation} 
U=\exp \left[ \frac{-it\eta ^{2}\Omega ^{2}}{\Delta }(\sigma _{x}^{i}\sigma 
_{x}^{j}+\sigma _{y}^{i}\sigma _{y}^{j})\right] =\exp (-itH^{ij}).
\end{equation} 
$H^{ij}$ is an example of an anisotropic Heisenberg exchange interaction (often referred to as the XY Hamiltonian). 
Previous work in our group \cite{DiVincenzo:00a,Kempe:01a,Kempe:01b,Kempe:01c,Vala:02b} has shown that such two-body exchange Hamiltonians can be utilized to perform universal quantum computation on encoded qubits. 
%We now show how this may be used to make quantum computation operations in an ion trap. 

\section{Encoded Universality in Ion Traps} \label{encoded} 
 
\subsection{Universality of Anisotropic Exchange Interaction} 
 
It has recently been shown by Kempe et al. \cite{Kempe:01b,Kempe:01c} that the anisotropic exchange (XY) interaction is sufficient to generate universal encoded single-qubit and two-qubit operations. 
In particular, these universal gates may be generated over encoded subspaces spanned by either of the following qutrit code states consisting of three physical qubits: $C_I = \{|001\rangle,|010\rangle,|100\rangle\}$ or $C_{II} = \{|110\rangle,|101\rangle,|011\rangle\}$.  
It has been shown that the anisotropic exchange interaction between two physical qubits within either of these encodings generates the full $su(3)$ algebra, and hence also the $SU(3)$ group, over the logical qutrits. 
Since $SU(2) \subset SU(3)$, selecting a qubit out of the 3 qutrit states results in universal single qubit operations, as long as the full $su(3)$ algebra is preserved.  
In addition, encoded two-qubit operations have been shown to be efficiently generated from the anisotropic exchange interaction.\cite{Kempe:01c,Vala:02b}. 
 
\subsection{Encoded Space is a Decoherence Free Subspace} 
 
The encoded space we employ here is a Decoherence Free Subspace (DFS) with respect to collective dephasing.
An operator corresponding to this error model is defined as $S_z = \sum_{k=1}^n \sigma_z^k$, reflecting that
all physical qubits of the computer simultaneously gather an identical random shift in the relative phase between their ground and excited level.
We illustrate the DFS on a simple example of a logical qubit encoded into the states $|01\rangle$ and $|10\rangle$ of two physical qubits. 
Collective dephasing simultaneously shifts relative phase of each of the physical two-level systems as follows: $|0\rangle \to |0\rangle $, $|1\rangle \to e^{i \phi }|1\rangle $.
The logical qubit stays protected against collective dephasing because this phase shift becomes just an unimportant global phase identical for both logical qubit states.
An important source of collective dephasing error in linear ion traps is long wavelength fluctuations in the ambient magnetic fields.  
Work by Kielpinski et. al \cite{Kielpinski:01,Kielpinski:02} shows that using the two qubit collective DFS states defined to protect against collective dephasing leads to an increased fidelity limited now only by the heating of the ion motional state.  
However the results of Kempe et al. \cite{Kempe:01b,Kempe:01c} have shown that the anisotropic exchange interaction requires at least {\it three qubits} to serve as the logical qubit, with two posible qutrit codes ($C_I$ and $C_{II}$).  
Therefore, as indicated above, one needs to add a third ion in order to form a single logical qubit in the minimal encoding for the pairwise XY interaction.  
 
The XY Hamiltonian $H^{ij}$ preserves the DFS and the qutrit codes. 
Within the DFS, the collective dephasing operator $S_z$  commutes with the operator algebra generated by the XY interaction \cite{Kempe:01a,Kempe:01b}.  
Intuitively, all states from either codes are transformed identically if all their physical qubits are simultaneously phase shifted as described in the example above. 
The qutrit codes $C_I$ and $C_{II}$ preserve the total number of qubits in the excited level \cite{Vala:02b}. Consequently, they are invariant under collective dephasing, $S_z$.  

The same properties with respect to the collective dephasing model are also valid for qubit encodings derived from qutrits \cite{Kempe:01c}. 
We illustrate this here on an example of the logical qubit encoded into the states $|001\rangle$ and $|010\rangle$. 
The relative phase between the code-words remains unchanged if all the physical qubits of value $1$ are simultaneously shifted by an arbitrary phase kick $e^{i\phi}$ in the course of the collective dephasing process \cite{Kielpinski:01}. 
 
In the present case, we implement the XY interaction in steps, alternating the   
$\sigma_x^i \sigma_x^j$ and $\sigma_y^i \sigma_y^j$ operations.  
During this alternation, we could fall out of the DFS because the interactions $\sigma_x^i \sigma_x^j$ and $\sigma_y^i \sigma_y^j$  do not commute with the collective dephasing operator, $S_z$.   
This departure from the DFS can however be minimized by rapidly switching between the $\sigma_x^i \sigma_x^j$ and $\sigma_y^i \sigma_y^j$ Hamiltonians.  This method of avoiding decoherence is similar to proposed bang-bang schemes \cite{Viola:98}.  
To illustrate this mechanism, we make a simple argument here to show why this is the case. 
 
Assume we would like to apply the Hamiltonian $H^{ij}$ for a time T. One could imagine first applying $\sigma_x^i \sigma_x^j$ for a time T then $\sigma_y^i \sigma_y^j$ for a time T.  Or one could apply both operators $n$ times, each for a time T/n: 
\begin{equation} 
\hat{U} = e^{-i (\sigma_x^i \sigma_x^j + \sigma_y^i \sigma_y^j) T} = (e^{-i \sigma_x^i \sigma_x^j T/n} e^{-i \sigma_y^i \sigma_y^j T/n})^n  
\end{equation} 
 
Assuming only collective decoherence, a simple population model, described below, shows a radically improved decoherence time. 
Assume that the states outside of the DFS decay with some rate $\gamma$. 
We expect that the total decoherence will be proportional to the integral over time of $\gamma$ times the population outside of the DFS, $P(t)$.  
An overestimate would be to assume no decoherence and to simply integrate over the population outside of the DFS during our operation.  
 
Let us examine the unitary operator $\exp [i\sigma^i_y\sigma^j_y t] = \cos (t){\bf I} + i\sin (t)\sigma^i_y\sigma^j_y$.  
If this operator was applied to a state in the DFS subspace, i.e. a qubit from either $C_I$ or $C_{II}$, the identity would of course preserve this codespace and the population that has left the codespace, $p(t)$, would then be proportional to $\sin^2 (t)$. 
The integrated population from 0 to $t$ would then be proportional to $\frac{t}{2} - \frac{1}{4} \sin (2t)$.  
Following this operator by the application of  $\exp [i\sigma^i_x\sigma^j_x t]$ will return us completely to the subspace, and by symmetry leads to a total integrated population proportional to $P(t)=t-\frac{1}{2}\sin (2t)$. 
 
Returning to Eq. [4], we see that by applying our operators for a time $T/n$, we can achieve a total integrated population proportional to $n P(T/n) = T - \frac{n}{2}\sin{2T/n}$. For $n$ large compared to $T$, the decoherence is simply proportional to $\frac{T^3}{3n^2}$.  
Therefore, one expects a quadratic improvement in the fidelity with the number of steps, $n$, Fig. [1]. 
\begin{figure}[t!]
\includegraphics[width=1.0\hsize]{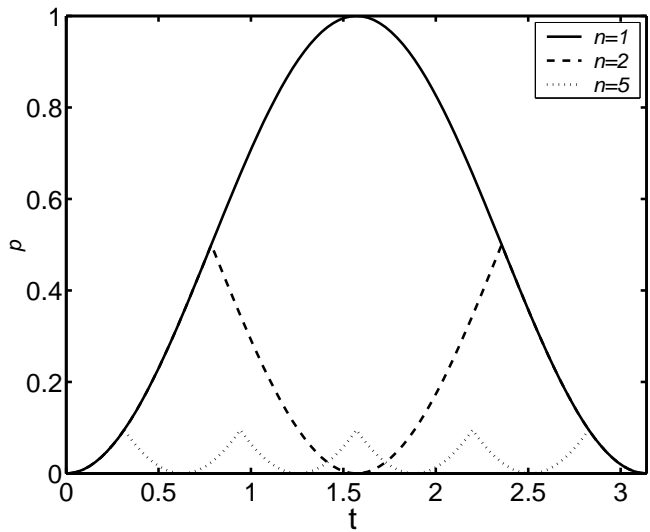}
\caption{A graphical demonstration of the difference in the population outside the DFS, $p(t)$, obtained by using a series of 2 $n$ alternating pulses, from that obtained by applying the $\sigma_x\sigma_x$ interaction followed by the $\sigma_y\sigma_y$ interaction once ($n$=1).  One can clearly see that the total integrated population outside of the DFS, $\int P(t)dt$, will be reduced as one increases the number of alternations, $n$.}
\end{figure}

To test this simple model, we calculated the evolution of the density matrix assuming collective decoherence and a Hamiltonian that alternates $n$ times between  $\sigma_x^i \sigma_x^j$ and $\sigma_y^i \sigma_y^j$. 
We then evaluated the fidelity, $f=tr \sqrt{\sqrt{\rho_a}\rho_b\sqrt{\rho_a}}$, between the evolved density matrix, $\rho_b$, and the expected density matrix resulting from direct application of the XY Hamiltonian, $\rho_a$. Fig. [2] shows that a fit of the operation error, $1-f$, on a $\pi$ pulse as a function of $n$ shows a quadratic improvement, as predicted by our simple population model.  
\begin{figure}[t!]
\includegraphics[width=1.0\hsize]{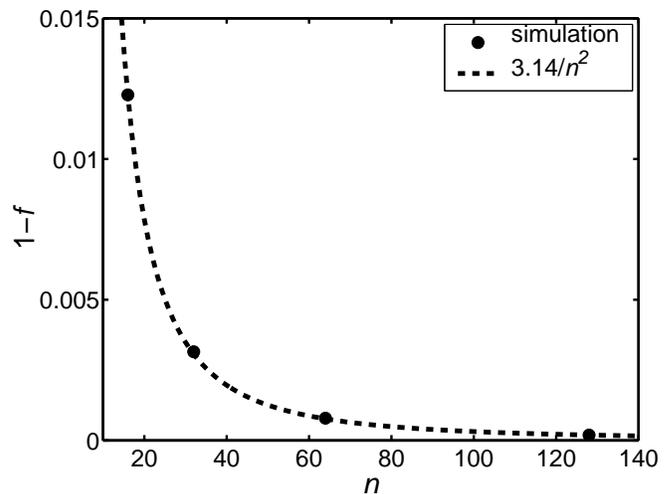}
\caption{The evolution of the density matrix for a single encoded qubit, consisting of three physical qubits ($C_I$ or $C_{II}$), was calculated using the Lindblad equation \cite{Lindblad:76} and assuming only collective dephasing.  The resulting system density matrix $\rho_a$ was then compared with the expected density matrix, $\rho_b$, if instead one could apply the XY Hamiltonian directly.
The plot shows operation error, 1-$f$, as a function of the number of steps, $n$, where $f$ is the fidelity, $f=\sqrt{\sqrt{\rho_b}\rho_a\sqrt{\rho_b}}.$  The strength of the collective decoherence, $\gamma$, was set equal to the strength of the driving Hamiltonian, $\frac{\eta^2\Omega^2}{\Delta}$, in this example.
The fidelity was calculated after a time $\frac{\pi\Delta}{\eta^2\Omega^2}$ .  A fit of the data for $n> 16$ yields the expected $1/n^2$ dependence (dashed line) predicted by our simple population model (see text).}
\label{fig2}
\end{figure}

\section{Conclusion} \label{end} 
 
We have presented a scheme for a universal ion trap quantum computation on Decoherence Free Subspaces using the Sorensen-Molmer two-body Hamiltonian.  
Rapid alternation of control pulses generates an effective  two-body interaction which approximates the anisotropic exchange (XY) Hamiltonian.  
This ensures universal quantum computation over DFS protected qubits, as a result of the anisotropic exchange interaction universality properties over logical qubits encoded into three physical qubits \cite{Kempe:01b,Kempe:01c,Vala:02b}.  
The decoherence time improves quadratically with the number of alterations, $n$, for a given total pulse duration $T$. 

Our scheme is scalable when one assumes the array-based architecture proposed by Kielpinski et. al. \cite{Kielpinski:02}. In that scheme single encoded qubits are held in individual traps.  Pairs of encoded qubits are then moved into the same trap in order to perform two qubit operations. The advantage of such a scheme
 is that one can avoid the unfavorable scaling of the two qubit interaction strength as one puts more ions into the same trap.  Furthermore, by encoding the ions into DFS, qubit coherences are protected when moving the ions from trap to trap.  

There are several differences for the physical implementation of our scheme compared to \cite{Kielpinski:02}. 
Our scheme requires one extra ion per logical qubit and each ion has to be individually addressed.  
Our operations take us out of the DFS, although this can be mitigated by the rapid alternation of pulses.  The alternation of pulses results in only a linear increase in the total number of physical operations. 
An advantage of our scheme is that we use only two qubit entangling operations, whereas the proposal of Kielpinski et al. \cite{Kielpinski:02} requires four-qubit entangling operations.
Two-qubit operations are inherently more robust to noise than four qubit entangling operations \cite{Sackett:00}.  
One could imagine a hybrid proposal in which the qubits are stored in pairs of ions.  
Single qubit operations are performed using the standard scheme \cite{Cirac:95,Kielpinski:01,Kielpinski:02}.  
Then when the qubits are brought together to do two qubit gates, two additional ions are placed in the trap allowing one to entangle the qubits only using two physical qubit operations as described here.  
This hybrid approach may offer greater flexibility.

\section{Acknowledgements}
The authors' effort was sponsored by the Defense
Advanced Research Projects Agency (DARPA) and Air Force Laboratory, Air
Force Materiel Command, USAF, under agreement number F30602-01-2-0524.
The work of KRB was also supported by the Fannie and John Hertz Foundation.

\end{document}